\documentclass[twocolumn,aps]{revtex4-1}

\usepackage{amsmath, amssymb}

\begin{document}

\title{The End of Eternal Inflation}

\author{Laura Mersini-Houghton and Malcolm J Perry}

\affiliation{DAMTP, University of Cambridge, Wilberforce Rd., Cambridge, CB3 0WA, England \\ and Department of Physics and Astronomy, UNC Chapel Hill, NC 27599, USA.}

\date{\today}

\begin{abstract}
We propose a new measure for eternal inflation that includes both conditions, large field fluctuations and smooth homogeneous domains, in the self reproducing probability estimate. We show that due to the increasing inhomogeneities in the background spacetime fractal, self-reproductions stops within a finite time $t_f$, thus inflation can not be eternal.
\end{abstract}

\maketitle
 
\section{Introduction}

Inflation is generally believed to be a natural way of solving three classic problems in 
cosmology; the flatness problem, the horizon problem and the monopole problem. It therefore
purports to explain why we live in state that is well described by a smooth Friedmann-Lemaitre-Roberstson-Walker ({\it FLRW}) universe. There is a difficulty with this scenario
and that is in the nature of the initial conditions that give rise to such inflation. Suppose that inflation is due to a single scalar field $\phi$ governed by a potential $V(\phi)$. Then in order 
to account for the observed present homogeneity of the universe, at the beginning of inflation 
the universe must be homogeneous on a scale about $10^4$ times the horizon scale. Thus
inflation has merely transformed one homogeneity problem into another one, as has been emphasised by Trodden and Vachaspati ~\cite{troddenvachaspati}.
In other words, the inflationary universe scenario requires an incredibly homogeneous initial state. The requirement of such a special initial state is the fine-tuning problem of inflation; or to 
put in a slightly different way, this requirement gives rise to the puzzle of why the pre-inflationary state should have such low entropy. The improbable initial state needed for starting inflation occupies a very small volume in the phase space of initial conditions.

However once inflation starts, quantum fluctuations of the metric and of the inflaton field $\phi$ with potential $V(\phi)$, generate density perturbations $\delta\rho / \rho$ with a scale invariant spectrum. The dynamics of large wavelength fluctuations is driven not only by the potential drift $\frac{\partial V}{\partial \phi}$ but also by the backreaction of short wavelength modes which provide a diffusion term $f(t,x)$ as described in more detail in the next section. When the dynamics of long wavelength perturbations is dominated by diffusion it leads to a stochastic behavior of these modes. The common lore is that a random large fluctuation will give rise to a newly produced inflating region, a new 'bubble' universe. Since the field undergoes Brownian motion in the diffusion regime then it enters a regime where large fluctuations are favored therefore the universe keeps reproducing {\it ad infinitum}  ~\cite{starobinsky, lindeselfreproduce, vilenkin, guth}. The self-reproducing regime is known as eternal inflation and it was first discovered by Linde in \cite{lindeselfreproduce}. The global geometry of the eternally inflating universe is highly nontrivial, it is a fractal where regions of large field excursions occupy a space volume with dimension $d<3$ ~\cite{vilenkinfractal, guth}.

Unlike scenarios of single shot inflation that are obtained by the extreme fine tuning of the initial state described above, eternal inflation appears to be unavoidable with many regions
undergoing inflation and to a picture of a universe filled with many bubbles that 
are the regions of spacetime that can be considered to be their own universes.
It is generic in the sense that all episodes of inflation arise spontaneously, without any fine- tuning of initial conditions and such events continue to occur for ever.

 The 'generic' reproduction of new universes seems to lead to an unbounded growth of phase space; in other words an infinite number of 'free lunches' is directly responsible for appearing to violate  unitarity on global scales. Eternal inflation leads then to a collection of difficult problems. Firstly,  spacetime is incomplete in the past so that there must still be some kind of an initial singularity ~\cite{borde, ggv}. Space has become fractal in nature with non-inflating regions being the fractal set. The measure for the geometry of space become non-normalizable 
associated to the apparent loss of unitarity. Tracing back in time this system of unbounded entropy and measure with a fractal space dimension, makes the problem of the improbable initial state an even more exquisitely special choice ~\cite{carrolltam, tegmark} .

In this letter we examine the basics of eternal inflation by addressing the question: what is the probability that random large fluctuations of a scalar field can lead to the self-reproducing inflationary regime?
We address this issue by taking into account the two ingredients needed for producing an inflationary bubble, namely the probability that the field will have a large fluctuation and also the probability that this large fluctuation will arise on a highly homogeneous patch of spacetime.   
As we show next, the probability to get a new bubble universe is 
\begin{equation}
P_i = P_{st} \times  P_{\phi}
\label{newprobability}
\end{equation}
where $P_{\phi}$ is the probability that the field has a large fluctuation and $P_{st}$ is the probability that this fluctuation does not \lq fall\rq\  in an inhomegenous small domain of the background spacetime. We show that this measure is not infinite, and it does not violate unitarity.  Further we show that this measure leads to the conclusion that producing new inflationary pockets, $P_i$ , is even more unlikely than just simply obtaining a random large fluctuation of the scalar field  $P_{\phi}$ , namely $\frac{P_i}{P_{\phi}} \ll 1$ since inhomogeneities from density perturbations grow quite fast thereby making the global spacetime geometry much more inhomogeneous in the future than the geometry at $t=0$ in the far past when inflation was not eternal. As we show below, generically $P_{st} \ll 1$.

\section{Basics and Setup}

Let us assume that we have an inflaton field $\phi$ in some slow roll potential, $V(\phi)$  in a {\it FLRW} universe with scale factor $a(t)$. In the absence of any spatial dependence of $\phi$, the field evolution is governed by
\begin{equation}
\ddot{\phi} + 3 H \dot{\phi} + \frac{d V(\phi)}{d \phi}  = 0   \, 
\label{fieldeqn}
\end{equation}
where
\begin{equation}
H^2 + \frac{k}{a^2} = \frac{8\pi G}{3} \left[ \frac{\dot{\phi}^2}{2} + V(\phi) \right]  \, 
\end{equation}
with $k = 0, \pm 1$ for flat, closed or open universes.

Inflation generates perturbations $\delta\phi$ which are frozen on superhorizon wavelengths with an amplitude $\delta\phi \simeq \frac{H}{2 \pi}$. Coarse-graining of the subhorizon fluctuations leads to a diffusion source term $f(x,t)$ for the long wavelengths modes \cite{starobinsky} such that
\begin{equation}
 < f(x_1 t_1) f(x_2 t_2) > = \frac{H^3}{4 \pi^2} \delta(t_1 - t_2) \frac{\sin z}{z}  \, 
\label{source}
\end{equation}
where $z = a H |x_1 - x_2|$. The coarse-graining of short wavelength modes results in a Langevin equation for the field \cite{starobinsky, lindeselfreproduce}
\begin{equation}
 3 H \dot{\phi}(x, t)   +  V'(\phi) = f(x, t)  \, 
\label{langevin}
\end{equation}
where the second derivative terms have, as usual, been ignored.
The probability distribution function for the field, $P_\phi$, satisfying Eq.~\ref{langevin} is given by a Fokker-Planck equation which provides the diffusion equation that describes the brownian motion of the field

\begin{eqnarray}
\frac{\partial P_{\phi}}{\partial t} = \frac{1}{3 H} \frac{\partial}{\partial \phi} \left[ V'(\phi) P_{\phi} \right]  + \frac{H^3}{8\pi^2} \frac{\partial^2 P_{\phi}}{\partial^2 \phi} \nonumber \\= \frac{\partial}{\partial \phi} \left[ \frac{1}{3 H} P_{\phi} \frac{\partial V}{\partial \phi} + \frac{\partial}{\partial \phi}  \left (D P_{\phi} \right)  \right]  \, .
\label{diffusion}
\end{eqnarray}
as was first described by Starobinsky ~\cite{starobinsky}.
The diffusion coefficient obtained after coarse-graining is $D = \frac{H^3}{8 \pi^2}$.
The first term in Eq. ~\ref{diffusion} gives the probability current due to the drift from the potential, while the second term gives the current due to the diffusion \cite{starobinsky, lindeselfreproduce, vilenkin}. If $V(\phi)=\frac{1}{2}m^2\phi^2$, corresponding to a free field of mass $m$, a stationary solution to Eq.~\ref{diffusion} is the same as the Hartle-Hawking (HH) expression for the probability, $P_{\phi} \simeq e^{-8\pi \frac{V(\phi)}{3 H^4}} \simeq e^{- \frac{(\phi - \phi_{0})^2 m^2}{6HD}}$ while the non-stationary solutions are more complex, as is discussed in detail, for example, in \cite{goncharev, starobinsky, yokoyama}. 

\section{A critical examination of the probability of eternal inflation}

\subsection{The Measure Problem}

Since $< \phi^2> \simeq D t$, then clearly, at large $t$ the diffusion term dominates over the drift term in Eq. ~\ref{diffusion}. Thus the field enters a regime where large fluctuations are favored. These estimates correspond to comoving volumes of horizon size $H^{-1}$ with a fixed comoving Planck length. The latter choice is useful in avoiding issues associated with transplanckian physics since a fixed physical Planck length leads to a  time-dependent phase space constantly replenished by transplanckian modes.

So far the procedure appears quite straightforward: one implements a coarse-graining scheme by integrating out the environmental degrees of freedom and finds out that the \lq bath\rq\ leads to a diffusion source term in the evolution equation for the \lq system\rq\ which sets the field into a random walk. In this case the \lq bath\rq\ is comprised of subhorizon wavelength modes and the \lq system\rq\ corresponds to long wavelength modes of the field. The combined dynamics of the drift and diffusion currents in Eq.~\ref{diffusion} determine the probability distribution function (PDF) for the field $\phi$ to be found at some sufficiently large value $\phi_{*} \simeq \phi + \delta\phi$ for a new \lq bubble\rq\ universe. Applied to a false vacuum decay potential this PDF is nothing more than the nucleation rate of bubbles $\Gamma \simeq e^{- S_4} $ with a $4-$action $S_4$ , that is the number of bubbles per unit $4-$volume. Applied to new, chaotic slow roll potential, or any other inflationary model, the PDF gives the {\it concentration of all the field fluctuations that reach the point $\phi_{*} \simeq \phi + \delta\phi$, i.e. the number of large fluctuations $N(\phi)$ that reach the point $\phi_{*} $ per unit spacetime volume}, that is 

\begin{equation}
 P_{\phi}  =  \frac{N(\phi_{*})}{V_{4}}  \,  . 
\label{oldprobability}
\end{equation}

where the time interval $\delta t = C H^{-1}$, (with $C<1$ a constant depending on the details of the potential), corresponds to intervals during which the field within a Hubble volume remains nearly unchanged.
At this stage, the conventional estimate that leads to the conclusion that inflation is eternal is based on the following two assumptions: i) it assumes that a large fluctuation $\phi_{*}$ will automatically guarantee the production of a new inflationary universe; ii) it assumes that the field PDF should be multiplied by the newly produced inflating volumes in physical coordinates, namely 

\begin{equation}
P_i  =  P_{\phi} V_{4}^{new} =C H^{-1} P_{\phi} V_{3}^{new} \,  .
\label{unboundold}
\end{equation}

 where \lq i\rq\ counts the newly produced bubbles.

We believe that assumptions $(i) + (ii)$ above are incorrect and they lead to the paradox of infinite measure.  The obvious conclusion obtained by following $ (i)+(ii)$ thus Eq. ~\ref{unboundold} , which ignores the spacetime homogeneity requirement on the estimates for eternal inflation,  is the result that the probability of bubble production, and thus of eternal inflation, can become not only large but in fact infinite due to the $4-$volume scaling in Eq.~\ref{oldprobability}. The main reason for the paradoxical conclusions, such as the $ill-$famous infinite measure problem, derived from these assumptions, is the fact that a large fluctuation does not necessarily produce a new universe. The unbound growth of phase space resulting from the conventional estimate of probability becomes apparent since Eq.~\ref{unboundold} introduces an extra source term, $3H(\phi) P_{\phi}$, in the diffusion equation Eq.~\ref{diffusion} written in physical coordinates, \cite{uzan}. The probability of the field fluctuation finding an homogeneous domain on the background spacetime is as crucial to the reproduction process as the energy of the field.  

We propose a new physical measure for the probability of eternal inflation $P_i$ instead of Eq.~\ref{oldprobability} by replacing $ (i) $ with a requirement of homogeneity on the initial state of any newly produced bubble for any potential inflationary domain and any fluctuation. We propose here that the PDF of the field $P_{\phi}$ should be multiplied  with the probability $P_{st} $ of finding a homogeneous domain in a highly inhomogeneous background in physical coordinates, produced by inflation, instead of multiplying it by the volume as previously done in Eq.~\ref{oldprobability}. 
As is well known having a large field fluctuation is a necessary but not sufficient condition for starting inflation since the spacetime region where this fluctuation arises should be exquisitely fine tuned to be smooth on scales larger than the Hubble volume ~\cite{trodden}. This is the unpopular fine tuning problem of inflation. Therefore, it does not matter how many large fluctuations we have, they will not give rise to inflationary pockets if they arise from inhomogeneous domains on the spacetime background. The two key issues relevant for the existence of inflation, instead of the number of large fluctuations $P_{\phi}$ and the physical volume ($ (i) + (ii)$), we demand are : 
$ a)$  the probability of finding a large fluctuation {\it and,} $ b)$ the probability that this large field excursion finds an homogeneous region in a highly inhomogeneous background. 

\subsection{Our proposal: The probability of eternal inflation}

Clearly demanding $(a)$ large field fluctuations and $(b)$ homogeneous spacetime regions for their excursions in estimating the probability of eternal inflation, leads to the expression of Eq.~\ref{newprobability} for the probability of inflation $P_i$, namely $P_i   =   P_{\phi} \times P_{st}$ 

The proposed probability in Eq.~\ref{newprobability} is our main point. Eq.~\ref{newprobability} provides the correct estimate for the probability of producing new inflationary regions, i.e. the measure of eternal inflation, since it takes into account both conditions that are needed for producing inflationary universe, namely large field fluctuations on smooth homogeneous spacetime domains. The reason why previous estimates of measures based on criterion $(ii)$ led to false results and infinite measures is because the estimated quantity from Eq.~\ref{oldprobability} is in fact the number of fluctuations passing through $\phi_{*}$ anywhere on the background space, and not the PDF for bubble production.
 Let us recall that the PDF for the field $P_{\phi}$ simply provides the concentrations of fluctuations with strength $\delta \phi$ that go through some space time point $(x,t)$ , 

\begin{equation}
P_{\phi} = \frac{N(\phi_{0} + \delta\phi)}{V_{4}}  \, .
\label{numberfluc}
\end{equation}

where $N$ is the number of all field excursions that start at $\phi_0$ and fluctuate to the point $\phi_0 + \delta\phi$ and $V_4$ is the $4-$volume. If we were to multiply this expression with the $3-$volume of space in physical coordinates, as was done in previous estimates of the measure $(i) +(ii)$, then what we are really calculating is not a probability but a number flux, namely the number of all field fluctuations that go through the point $\phi_{*} = \phi_0 + \delta\phi$  at some time $t$ anywhere in the physical space $3-$volume $V_3$. Certainly, the number of fluctuations should be infinite, thus it is no surprise that the measure always ends up being infinite. The previous estimate of probability for eternal inflation based on assumptions $(i) + (ii)$ rather paradoxically leads to a growing phase space. These two assumptions taken together lead to a picture where one estimates the total {\it number} of fluctuations anywhere in the physical space time volume (which of course is infinite) and equates that with a probability distribution of inflating pockets, therefore a growing phase space of inflation. To illustrate this point further we can use the example of false vacuum decay: if we were interested in the total number of instanton bubbles anywhere in the spacetime volume, then the number of instantons would obviously be infinite even for the dilute gas case, since in physical coordinates the space volume is $V_3 = a^3 \simeq e^{3 Ht}$ and therefore according to Eq.~\ref{oldprobability} we should have that the probability of eternal inflation is $P_i \simeq H^{-1}\Gamma V_3 \simeq H^{-1} e^{- S_4 + 3 H t}\rightarrow \infty $.

Next, we would like to estimate the probability of large fluctuations finding homogeneous domains on the background spacetime, $P_{st}$ in order to investigate the results of the new measure Eq.~\ref{newprobability}. Generally in a highly inhomogeneous spacetime such as this, homogeneous regions are the inflating domains. Therefore, $P_{st}$ is given by the ratio of volumes occupied by inflating regions over the global volume of spacetime

\begin{equation}
P_{st}  =  \frac{V^h}{V^{tot}}  \, .
\label{homogenprob}
\end{equation}

\subsection{The probability of homogeneous domains $P_{st}$ }

A consistent treatment of eternal inflation should incorporate the relation of diffusive perturbations Eq.~\ref{diffusion} with the global spacetime metric via Einstein equations. Long wavelength inhomogeneities produce density perturbations $\frac{\delta\rho}{\rho}$ which grow at large scales. Thus the metric of spacetime on global scales is quite complex and highly inhomogeneous. Part of this structure can be captured by the following metric with a space and time dependent scale factor $a(x,t)$

\begin{equation}
ds^2 \simeq dt^2 - a(x, t)^2 dx^2 \,
\end{equation}

Let us emphasise the complex inhomogeneous structure of spacetime resulting from the dissipative field dynamics by reviewing some previous results obtained in ~\cite{goncharev}. These authors ~\cite{goncharev} used a conformally Newtonian coordinate system in order to relate the field to the metric

\begin{equation}
ds^2 \simeq \left(1 + 2 \Phi(x, \tau)\right)d\tau^2 - \left(1 - 2 \Psi(x,\tau)\right)dx^2 \,
\end{equation}
(with $\tau$ the usual conformal time), and show that the relation obtained by Einstein equations and perturbation equation, $\frac{\delta\rho}{\rho} \simeq - 2 \Phi$ and $\Psi \simeq \Phi$ holds, for at least around an inflating region. In this gauge, the relation between the background potential $\Phi$ and the field fluctuations leads in a straightforward manner to a diffusion equation for $\Phi$  which is analogous to Eq.~\ref{diffusion} for the field diffusion. That is, the metric of spacetime responds adequately to field excursions due to diffusion, as it should based on Einstein equations. The background potential $\Phi$ has a dispersion $\Delta_{\Phi}$ which also grows with time
\begin{equation}
\Delta_{\Phi}^{2} \simeq < \Phi^2 > \simeq \frac{2^{1/2} V'^2}{3 (3 \pi V M_{p}^{3} )^{1/2}}  \tau  \,
\label{metricdispersion}
\end{equation}

The growing dispersion on the metric is not surprising since we expect $\frac{\delta\rho}{\rho} \gg O(1)$ at large scales. Since $\frac{\delta\rho}{\rho} \simeq - 2 \Phi$ during inflation that naturally density perturbations grow with time with the same dispersion as $\Delta_{\Phi}^{2} $. 

The key issue in these estimates is the fact that globally the universe becomes highly inhomogeneous. Further work showed that in fact the global geometry of the universe is that of a fractal whereby homogeneous regions with $\frac{\delta\rho}{\rho} \ll O(1)$ occupy a very small volume of dimension less than three ~\cite{vilenkinfractal}. Since the spacetime metric tracks the field then the probability of finding homogeneous regions at large times, estimated in ~\cite{vilenkinfractal} is $P \simeq \int_{-\phi_0}^{\phi_0} P_{\phi} d\phi \approx \frac{4}{\pi} e^{-\frac{D \pi^2 t}{4\phi_{0}^{2}} }$ where $\phi_{0}$ is the boundary value of the field's excursion. This estimate leads to inhomogeneous spatial regions with fractal dimension $d$ having a volume 
\begin{equation}
V_{3}^{inh} \simeq e^{d H t}  \qquad d = 3 - \frac{D \pi^2}{4 \phi_{0}^2}  \,
\label{inhomogvolume}
\end{equation}
For the case of an effective potential $V = V_0 - \frac{m^2 \phi^2}{2}$ with $m^2 \ll H^2$ the fractal dimension is $d = 3 - \frac{m^2}{3 H^2}$ and the probability of not having inhomogeneous regions is $P \simeq e^{-\frac{m^2}{3 H^2} t}$. Similarly for $V = V_0 - \lambda \phi^4 /4 , d = 3 - \frac{\lambda^{1/2}}{3^{1/2} 4\pi}$. Starobinsky and Yokoyama ~\cite{yokoyama} studied the stochastic behaviour of the field on a DeSitter (DS) background and estimated the two point correlation functions in space and time and showed that the spatial correlation size does not depend on time, i.e. the physical size of domains  $h$ with growing and decreasing field values remains constant independent of the DS background expansion.

The implications of these results is that since the number of homogeneous and inhomogeneous domains, even on a DS background, scales with the expansion then we can use the scaling properties of fractals to estimate the probability that a large field fluctuation finds a homogeneous domain on this fractal background. The evolution of each $h-$domain with $|\phi| \le H$ is equivalent to the evolution of a single large volume $V_{3}^{tot}$ with size $R$ which started inflating at the initial time $t=0$ ~\cite{vilenkinfractal} since the stochastic field excursions and therefore the evolution of each $h-$domain is independent from each other. Therefore, the total spatial volume for all homogeneous domains of size $x \ll R$ in this background of size $R$ occupies a volume $V^h \simeq x^3 (\frac{R}{x})^d$ due to the self similar properties of the fractal geometry. It follows that the probability of finding an homogeneous region of size $x$ at any time 't' in this inhomogeneous background of size $R$ is 
\begin{equation}
P_{st} = \frac{V^{h}}{V^{tot}} = \left(\frac{x}{R} \right)^{(3 - d)}  \,
\label{homoprob}
\end{equation}

For example, for chaotic or new inflation type potentials $P_{st} \simeq (\frac{x}{R})^{\frac{m^2}{3 H^2}}$. Note that $R = H^{-1} \tilde{a(t)}$ where $\tilde{a(t)}$ is some averaged value of the background expansion scale factor. Then
\begin{equation}
P_{st} \simeq (H x)^{\frac{m^2}{3H^2}}  e^{-(\frac{m^2}{3 H^2}) H t} \,
\label{homoprobexample}
\end{equation}

This relation holds for any potential $V$ provided that the deviation $\alpha$ of the fractal dimension from the background space of dimension three, $\alpha = [3 -d]$ replaces $\alpha = m^2/3H^2$ in the above example.
It is clear from Eq.~\ref{homoprobexample} that inhomogeneities grow too much within a time $t_f  \simeq \frac{3H}{m^2} \simeq (\alpha H)^{-1}$. The critical timescale for the growth of inhomogeneities implies that the probability of a fluctuation finding an homogeneous region in order to produce an inflationary pocket, is incredibly small for all times $t \ge t_f$, according to our proposed measure.  This result should be expected of any fractal geometry since the volume of homogeneous regions has a dimension less than three thus cannot fill the $3-$volume of space and this homogeneous volume is increasingly diminishing with time relative to the background volume, as we proved in Eqs.~\ref{homoprob}, ~\ref{homoprobexample}. The resulting spacetime becomes highly inhomogeneous within a time interval, $0<t<t_f$, and the probability of finding smooth domains becomes smaller and smaller. Large fluctuations cannot give rise to inflationary pockets without arising on homogeneous domains. In short, inflation cannot self reproduce beyond the point $t>t_f$.

The full measure we proposed here, $P = P_{\phi} \times P_{st}$ for example in the case of chaotic type inflation where the field has to fluctuate upwards in the potential, yields the following probability for eternal inflation

\begin{equation}
P = P_{\phi} \times P_{st} \simeq B e^{- \frac{(\phi^2 - \phi_0^2)}{2 D}} (Hx)^{\frac{m^2}{3H^2}}  e^{-(\frac{m^2}{3H^2}) Ht}   \,
\label{fullprob}
\end{equation}
with $B$ a normalization constant.
The case of false vacuum decay potential leads to exactly the same result by replacing in Eq.~\ref{fullprob} the fluctuation PDF with the bubble nucleation rate $P_{\phi} = \Gamma$ and the fractal dimension $\frac{m^2}{3H^2} = [3 - d] $ with the result relevant to the false vacuum decay fractal $[3-d] = 3 - \frac{4\pi}{3} \Gamma$ obtained in ~\cite{ggv}.

\section{Conclusion}

We have shown here that inflation cannot sustain an eternal replenishment of 
\lq free lunches\rq\  of new universes after a time $t>t_f \simeq \frac{1}{(d-3)H}$ since the background space grows highly inhomogeneous. Therefore one of the two conditions for producing inflationary regions namely: $(a)$ a large fluctuation and $(b)$ an exquisitely homogeneous domain,  is not fulfilled after time $t>t_f$.

We proposed a new measure that accurately takes both requirements of producing inflation into account in the probability expression. All previous proposals did not include the homogeneity condition $(b)$ into the probability measure. The new measure is finite and normalizable, and it preserves unitarity. The latter can be seen by the fact that the diffusion PDF for the field is directly converted to a diffusion PDF for the entropy of fluctuations, $S \simeq log[P_{\phi}]$. Since individually large fluctuations spread in phase space due to diffusion and dissipation, then the volumes their trajectories occupy in the phase space grows. But the whole closed system that includes all the excursions does not change. The closed system occupies a volume in phase space bound by a finite entropy $S$ given by the background horizon $H^{-2}$. The trajectory of each individual fluctuation can only spread and grow within the finite phase space volume occupied by the closed system. In short,the phase space volume of the whole system and its finite entropy are larger than that of individual bubbles and fluctuations. Although the entropy of these subsystems can grow, it is finite and can only grow inside the bounds provided by the whole background system. The entropy of the background system is that of the field at the hypersurface of initial conditions, i.e the entropy at the initial moment $t=0$ when inflation first switches on which does not change.

The new measure we propose removes some of the problems and paradoxes related to previous measures of eternal inflation. Previous measures were unbounded because they resulted in large fluctuations being favored and equated that process with the production of new bubble universes. The probability of decaying to false vacuum or jumping higher up the potential were favored over the decay to lower energy states. Further, damping was missing from the Einstein-Smoluchowsky (E-S) type relation of dissipation to damping, due to the infinities produced. It is easy to see that with the new measure the E-S relation $D/\mu \simeq kT$, where $\mu$ is the damping coefficient, is satisfied since $T\simeq H^{-1}$ therefore $\mu \simeq D H$ is related to $\mu \simeq (\frac{D}{d-3})\frac{1}{t_f}$. 

Before the proposed measure here, we were in a situation where either probability was not conserved or, it was not finite, or both. Besides, the width of the wavepacket $<\delta\phi^2> \simeq Dt >> 1$ loses quantum coherence which makes the exponential expansion of the newly produced $h-$domains nearly impossible ~\cite{yokoyama}. Even fluctuations with $< \delta\phi^2 >  \ll 1$, run the danger of $<\delta p_{\phi}^2 > >1$ by the uncertainty principle in which case slow roll inflation can not be obtained despite large field fluctuations. Taking into account the homogeneity requirement precludes the existence of coherence for the wavepacket which ensures that we recover the correspondence with the usual semiclassical equations of inflation where coherent quantum fluctuations behave as classical states. We have shown here that, despite the diffusive behaviour of fluctuations, the probability of subsequent episodes of inflation is highly suppressed relative to even the probability of single shot inflation which is already incredibly small. Thus inflation can not be eternal.

{\it Acknowledgements : }
 LMH is grateful to Mark Trodden for useful discussions and to $DAMTP$, Cambridge University, for their hospitality during the time part of this work was done.
MJP would like to thank Gary Gibbons for discussions. LMH acknowledges the DOE support of grant DE-FG02-06ER1418 and of the Bahnson fund. MJP is in part supported by the STFC rolling grant $STJ000434/1$. MJP would like to thank the Mitchell family foundation and Trinity College Cambridge for their support.


\end{document}